\documentclass{article}

\usepackage{amsmath,amssymb}

\begin{document}
\title{An exact universal amplitude ratio for percolation.}
\author{Katherine A. Seaton
\thanks{
C. N. Yang Institute for
Theoretical Physics, State University of
New York, Stony Brook NY 11794-3840,USA}
\thanks{
Permanent address:
School of Mathematical and Statistical Sciences,
La Trobe University, Victoria 3086, Australia}
\thanks{email: k.seaton@latrobe.edu.au}}

\maketitle

\begin{abstract}The universal amplitude ratio $\tilde{R}_{\xi}^{+}$ for
percolation in two dimensions is determined exactly
using results for the dilute A
model in regime 1, by way of a relationship with the $q$-state Potts
model, with $q\leqslant 4$. 
\end{abstract}


Percolation is a subject which can be described simply and 
yet has many important physical applications (see \cite{SA}
and references therein). In statistical mechanics, one natural
arena in which to study its inherent critical phenomena,  
percolation is related by way of the random cluster model
\cite{FK} to  the $q$-state Potts model with $q=1$. 

The {\it critical exponents}
of the two-dimensional Potts model 
for $q\leqslant 4$, determined from numerical and
renomalization group studies \cite{dn, nien},
led to the identification \cite{dot}  of  the Potts model 
(in the scaling limit) 
for $q=2,\, 3,\, 4$ 
with the $\phi_{2,1}$
perturbation of the minimal unitary series $\mathcal{M}(t,t+1)$
by way of
\begin{equation}
\sqrt{q}=2 \sin\left(\frac{\pi(t-1)}{2(t+1)}\right). \label{qt}
\end{equation}
With care, one can extend the identification to $q=1$.

On the other hand, for the
associated {\it critical amplitudes} (except in the Ising case $q=2$
where exact results are known \cite{McWu}) the series and Monte Carlo
results can vary widely \cite{wu}, making it difficult to obtain
reliable estimates for the universal amplitude ratios \cite{PHA}.
Recently numerical values for 
various amplitude ratios of the Potts model (including percolation) have
been determined \cite{dc} by a technique of perturbed conformal field
theory, specifically by the two-kink approximation in the
form factor approach  to the $S$-matrix of \cite{chim}.

The only one of the amplitude ratios determined in \cite{dc}
which will be considered in this letter is the percolation
amplitude \cite{PHA, dc}
\begin{equation}
\tilde{R}_{\xi}^{+}=
\lim_{q\to 1}\left[\frac{\alpha(2-\alpha)(1-\alpha) \mathcal{A}_f}{(q-1)}
\right]^{\frac{1}{2}}\xi_{0}^{+}. \label{Imp}
\end{equation}
Here 
$\xi_0^{+}$ is the leading term amplitude of the correlation length
(above the critical temperature)
\begin{equation*}
\xi \simeq \xi_0^{+} \tau^{-\nu},
\end{equation*}
and $\mathcal{A}_f$ and $\alpha$ come from the singular 
part of the free energy, 
\begin{equation*}
-f_{\rm s} \simeq {\mathcal
A}_f \tau^{2-\alpha}.
\end{equation*}
The percolation analogue of the free energy is the mean number of clusters, 
and of
$\xi$, the pair connectivity.
That $\tilde{R}_{\xi}$ is universal, i.e. independent of metric factors 
associated
with the reduced temperature $\tau \propto T-T_c$,  follows
from the scaling relation $2-\alpha =d \nu$ in $d=2$ dimensions.

There is an exactly solvable model, the dilute A$_L$ model
\cite{WNS}, which has also been identified with the
$\phi_{2,1}$ perturbation of the minimal unitary series \cite{WPSN}.
It is an $L$ state, 
interaction-round-a-face model, solved in four regimes,
two of which provide off-critical extensions of the unitary minimal conformal
field theories \cite{WPSN}.  While regime 2 of the model is associated with
the perturbation $\phi_{1,2}$, regime 1 of the model realizes
(in the scaling limit) the perturbation $\phi_{2,1}$ of the minimal
unitary series $\mathcal{M}(L+1,L+2)$.

That they are both identified with the same perturbation $\phi_{2,1}$ 
suggests a relationship at least
between certain members of the dilute A$_L$ hierarchy and the Potts models. 
In general the Potts model and the dilute A model represent 
different universality classes associated with $\phi_{2,1}$, just
as in the field theory context there may be more than one $S$-matrix 
for a given perturbation. 
One simple way to see that the models have different internal 
symmetries is to count ground states: for the Potts
model there are $q$ of them, and for the dilute A model the number
grows linearly with $L$, but the connection between $L$ and $q$
implied by setting $t=L+1$ in (\ref{qt}) is certainly not linear.

Nevertheless, on the basis of this relationship between models,
weaker than shared
universality class, it is still possible to determine one
universal amplitude ratio for the Potts model, including 
percolation, from the dilute
A model. 
In the language of perturbed conformal field theory, the free energy and 
the correlation
length are related directly to the coupling constant $g$ of the perturbation 
and the
associated conformal weight $\Delta$:
\begin{equation*}
f_{\rm s} \sim  g^{\frac{d}{d-2\Delta}} \qquad \qquad
\xi    \sim  g^{\frac{-1}{d-2\Delta}}.
\end{equation*}
Thus when attention is confined to the correlation length amplitude ratio 
(\ref{Imp}),
the required quantities $\mathcal{A}_{f}$, $\xi_0$ and $\alpha$ (or 
equivalently $\Delta)$
relate solely to the perturbing operator. This operator is $\phi_{2,1}$
for both  dilute A in regime 1 and the Potts model, and any universal 
observable associated only
to it should be common \cite{delf2}.

From (\ref{qt}), percolation corresponds to $t=2$, so the member of
the dilute A hierarchy to be considered is $L=1$.
Since $L$ labels the number of states in the dilute A$_L$ model, 
it is properly an integer $L\geqslant 2$.
However, it has long been realized that $q$ can
be treated as a continuous variable in an appropriate
formulation of the Potts model \cite{FK} and it is in this spirit
that we now take $L \to 1$ in the free energy and correlation length
expressions of the dilute A model.

The leading term of the correlation length \cite{BS, BS2} of the dilute A$_L$
model in regime 1 is
\begin{equation*}
\xi^{-1}\simeq 4 \sqrt{3}\  p^{2(L+1)/3L}. 
\end{equation*}
There is good reason to believe that this 
applies to the high temperature regime for all $L$, although it
was actually determined from $L$ odd alone.

Although the singular part of the free energy of the dilute A model has
been determined both by the inversion relation \cite{WPSN}
and exact perturbative \cite{BS,BS2} approaches, the coefficient
has not previously been explicitly given; apart from $L=2$, in regime 1
it may be written
\begin{equation}
-f_{\rm s}\simeq\frac{4 \sqrt{3} \sin(2 \pi(L-1)/3L)}{\sin(\pi(L-2)/3L)}
p^{4(L+1)/3L}.
\label{free}
\end{equation}
This correctly gives the critical exponent $\alpha=-2/3$ for percolation
by setting $L=1$ in
\begin{equation*}
\alpha= 2-\frac{4(L+1)}{3L}=\frac{2(L-2)}{3L}. \label{alf}
\end{equation*}
The coefficient in (\ref{free}) vanishes at $L=1$, which agrees
with the $q=1$ Potts model amplitude given by Kaufman and Andelman \cite{KA}.
 
The universal amplitude
ratio $\tilde{R}^{+}_{\xi}$, however, is finite and non-zero. 
Using trigonometric identities in (\ref{qt}) it is possible to write
\begin{equation*}
q-1=4\sin\left(\frac{\pi(2t-1)}{3(t+1)}\right)
\sin\left(\frac{\pi(t-2)}{3(t+1)}\right),\label{rule}
\end{equation*}
in which we now set $t=L+1$.
Using the last four expressions
in (\ref{Imp}), and taking the limit $L \to 1$ gives the algebraic
expression  for percolation:
\begin{equation}
\tilde{R}^{+}_{\xi}=\left[\frac{(L-2)(L+1)(L+4)(L+2)}{27 \sqrt{3}L^4
\sin\left(\frac{\pi (2L+1)}{3(L+2)}\right)\sin\left(\frac{\pi(L-2)}{3L}
\right)}
\right]^{\frac{1}{2}}_{L=1}=\left[\frac{40}{27\sqrt{3}}\right]^{\frac{1}{2}}.
\label{exact}
\end{equation}
To compare this result with the numerical field theory result of 
\cite{dc}, it is to four figures
\begin{equation*}
\tilde{R}_{\xi}^{+}=2^{\frac{3}{2}}3^{\frac{-7}{4}}5^{\frac{1}{2}}=
0.9248\ldots\ .
\end{equation*}

Delfino and Cardy \cite{dc} quote a previous best estimate 
$\tilde{R}_{\xi}^{+} \approx
1.1$ based on earlier series \cite{dp} and 
Monte Carlo \cite{MC}
results.  By the technique they adopt,
they obtain for comparison the numerical value $\tilde{R}_{\xi}^{+}=0.926$. 
Both of these values use the second moment correlation length
which differs (only slightly for $q=1$) from the true
correlation length used in this letter, though in \cite{dc}
the expression (\ref{exact}) could also be obtained. 
Numerical values for other universal amplitude 
ratios are also found in \cite{dc};
unlike (\ref{exact}),  these 
do not all appear to be accessible exactly or from existing solvable models. 
Thus the value of knowing the reliability
of the field theory approach, specifically the
form factor approach at the two kink level, is clear (see \cite{cm}).

A more detailed explanation of the relationship between the $q$-state
Potts model and the dilute A model, together with the calculation 
of the universal correlation length amplitude
ratio $R_{\xi}^{+}$ for the $q$-state Potts model for all integer $1\leqslant q
\leqslant 4$, will be given elsewhere \cite{kas2}.

\section*{ACKNOWLEDGMENTS}
This work was completed during sabbatical (OSP) leave
spent at the Department of Mathematics and Statistics, University
of Melbourne and at SUNY Stony Brook. The author particularly acknowledges
Barry McCoy's useful discussions and hospitality. She also thanks
Aldo Delfino for informative comments on the first version of this letter.


\end{document}